\documentclass[twocolumn,showpacs,preprintnumbers,amsmath,amssymb]{revtex4}
\usepackage{array}
\usepackage{booktabs}
\usepackage{tabu}
\usepackage{dcolumn}
\usepackage{amsmath}
\usepackage{amsfonts}
\usepackage{amssymb}
\usepackage{graphicx}
\usepackage{subfigure}
\usepackage{graphicx}% Include figure files
\usepackage{dcolumn}% Align table columns on decimal point
\usepackage{bm}% bold math

\def\be{\begin{equation}}
  \def\ee{\end{equation}}
\def\bea{\begin{eqnarray}}
\def\eea{\end{eqnarray}}
\def\f{\frac}
\def\n{\nonumber}
\def\l{\label}
\def\p{\phi}
\def\o{\over}
\def\R{\rho}
\def\pa{\partial}
\def\om{\omega}
\def\na{\nabla}
\def\P{\Phi}
\begin{document}

\title{Intermediate inflation driven by DBI scalar field}% Force line breaks with \\

\author{N. Nazavari}
% \email{t.golanbari@uok.ac.ir}
  \author{A. Mohammadi}
 \email{abolhassanm@gmail.com}
   \author{Z. Ossoulian}
% \email{t.golanbari@uok.ac.ir}
   \author{Kh. Saaidi}
% \email{ksaaidi@uok.ac.ir}

\affiliation{
Department of Physics, Faculty of Science, University of Kurdistan,  Sanandaj, Iran.\\}
\date{\today}% It is always \today, today,

\def\be{\begin{equation}}
  \def\ee{\end{equation}}
\def\bea{\begin{eqnarray}}
\def\eea{\end{eqnarray}}
\def\f{\frac}
\def\n{\nonumber}
\def\l{\label}
\def\p{\phi}
\def\o{\over}
\def\R{\rho}
\def\pa{\partial}
\def\om{\omega}
\def\na{\nabla}
\def\P{\Phi}
%\nofiles

%=============================================================%
%=============================================================%
%============== Abstract =======================================%
%=============================================================%
%=============================================================%
\begin{abstract}
Picking out DBI scalar field as inflation, the slow-rolling inflationary scenario is studied by attributing an exponential time function to scale factor; known as intermediate inflation. The perturbation parameters of the model are estimated numerically for two different cases and the final result is compared with Planck data. The diagram of tensor-to-scalar ratio $r$ versus scalar spectra index $n_s$ is illustrated, and it is found out that they are in acceptable range, as suggested by Planck. In addition, the acquired values for amplitude of scalar perturbation reveals the ability of the model for depicting a good picture of the universe in one of the earliest stage.  As a further argument, the non-Gaussianity is investigated displaying that the model prediction stands in $68\%$ CL regime; according to latest Planck data.
\end{abstract}
%Use showkeys class option if keyword
\maketitle

%%%%%%%%%%%%%%%%%%%%%%%%%%%%%%%%%%%%%%%%%%%%%%%%%%%%%%%%%%%%%%%%%%%%%%%%%%%%
%%%%%%%%%%%%%%%%%%%%%%%%%%%%%%%%%%%%%%%%%%%%%%%%%%%%%%%%%%%%%%%%%%%%%%%%%%%%
%%%%%%%%%%%%%%%%%%%%%%%%%%%%%%%%%%%%%%%%%%%%%%%%%%%%%%%%%%%%%%%%%%%%%%%%%%%%
%%%%%%%%%%%%%%%%%%%%%%%%%%%%%%%%%%%%%%%%%%%%%%%%%%%%%%%%%%%%%%%%%%%%%%%%%%%%
%============  Sec.I (Introduction)  =======================================
%%%%%%%%%%%%%%%%%%%%%%%%%%%%%%%%%%%%%%%%%%%%%%%%%%%%%%%%%%%%%%%%%%%%%%%%%%%%
%%%%%%%%%%%%%%%%%%%%%%%%%%%%%%%%%%%%%%%%%%%%%%%%%%%%%%%%%%%%%%%%%%%%%%%%%%%%
%%%%%%%%%%%%%%%%%%%%%%%%%%%%%%%%%%%%%%%%%%%%%%%%%%%%%%%%%%%%%%%%%%%%%%%%%%%%
%%%%%%%%%%%%%%%%%%%%%%%%%%%%%%%%%%%%%%%%%%%%%%%%%%%%%%%%%%%%%%%%%%%%%%%%%%%%
\section{Introduction}
Inflationary scenario is undoubtedly one of the most successful candidate to describe our Universe in the earliest stage, supported strongly by large number of observational data \cite{Linde}. Besides solving the three major problem of Hot Big-Bang theory, the scenario predicts two important type of perturbations as scalar and tensor perturbations. The scalar perturbations are seeds for large scale structure of the universe, and the tensor perturbations are known as gravitational waves as well \cite{Bardeen-etal,Mukhanov-etal,Malik}. \\
So far, there are many types of inflation proposed by cosmologists, however it could be said that the slow-rolling models of inflation are the most popular ones. The simplest model of inflation is driven by a minimal coupling scalar field, inflaton, with a self interacting potential govern by Einstein gravity; known as standard model of inflation. The model could be modified in different ways \cite{Baumann}. A simple and interesting generalization of the standard inflationary scenario is to extend the kinetic part of the inflaton \cite{Unnikrishnan}; known as k-inflation which their properties lead to huge interest amongst scientists \cite{k-inflation,saaidi,golanbari}. \\
General theory of relativity is believed to be a low energy effective field theory from string theory viewpoint, that undergoes some corrections in classical and quantum level \cite{Spalinski}. Up to now, wide kinds of scalar fields are anticipated by string theory related to compactification of extra dimensions and brane motion in a bulk \cite{Bessada}. Significant role of inflation in the modern cosmology leads physicists to study the scenario in string theory framework. Brane inflation is a possible scenario \cite{Dvali,Burgess,Kachru}. One of the suggested scalar field by string theory comes from Dirac-Born-Infeld action, known as DBI scalar field, so that a non-canonical kinetic term is attached to the scalar field. Then, DBI scalar field could be thought as another case of k-inflation.  \\
Studying slow-rolling inflationary scenario is performed in some different ways. The most common way amongst cosmologists is to put an specific form of potential, taken mostly from particle physics like Higgs potential, in dynamical equations and redefine the slow-rolling parameters in term of the potential. Introducing scale factor is another way to consider inflationary scenario, where the "intermediate inflation" is the most popular one; presented first by \cite{Barrow}. In this approach, an exponential function of time describes the scale factor behavior, $a(t) = \exp\big( A t^\alpha \big)$, $A>0$ and $0<\alpha<1$ \cite{Vallinotto,Starobinsky}. This can be attained from a potential asymptotically looks like negative power but not exactly \cite{Rendall}. The scenario indicates on a expansion faster than power-law inflation ($a(t)=t^p, p>1$), and slower than de-Sitter inflation ($a(t)=\exp(Ht), H=cte$). Intermediate inflation in Einstein gravity creates a scale invariant perturbation when $\alpha=2/3$ \cite{Barrow,Barrow-etal,Vallinotto,Starobinsky}. There are large number of literatures \cite{Muslimov,BarowLiddle02,BarrowLiddle,Rezazadeh,mohammadi} that widely have discussed the intermediate inflation, which in turn points out the fact that the scenario holds a suitable place in the scientists' project. The scenario is able to satisfy the bound on scalar spectra index $n_s$ and tensor-to-scalar ratio $r$, measured by observation on CMB \cite{BarrowNunes}. \\
Inflation driven by DBI scalar field has been investigated, where an specific form of the potential  was picked out. For instance, in \cite{Li}, the authors seeks to establish a relation for the observational parameter, where they found out that the case comes to some difficulty in matching with Planck data. The inflationary model of DBI for different classes of the potential is presented in \cite{Bessada}. Besides these, the case of power-law inflation for DBI scalar field is performed in \cite{Spalinski,Spalinski-a}. The presented work is proceeded by utilizing a different approach. Intermediate inflation with DBI scalar field in Einstein gravity is the main goal of the work which will be pursued in the following context.  \\
The paper is organised as following: a brief review of the DBI scalar field given in Sec.\ref{Sec2}, where the main dynamical equations of the model are presented. Intermediate inflation is introduced in Sec.\ref{Sec3}, where the main parameters are presented generally, especially those related to perturbations. It will be followed by Sec.\ref{Sec4}, where the final result of the model about scalar spectra index, tensor-to-scalar ratio, and amplitude of scalar perturbation are computed numerically, and they are compared with latest observational data given by Planck collaboration. Sec.V is associated to the energy scale of the inflation so that the initial energy scale of the universe is demonstrated to be smaller than Planck energy, where the quantum effects in the model could be ignored. The non-Gaussianity as another important observational parameter is studied in Sec.VI in which figures out to stay in $68\%$ CL regime. To sum up, a brief report of the work and the result is brought forward in conclusion section.    \\

%=============================================================%
%=============================================================%
%============== Section 2 =======================================%
%=============================================================%
%=============================================================%
\section{An overview of DBI}
\label{Sec2}
Firstly, we review the main dynamical equations in DBI scalar field in Einstein gravity (refer to \cite{Copeland,Spalinski-a} for more detail about the model). Consider the following action for DBI scalar field
\begin{equation}\label{1}
S=-\int d^{4} x \sqrt{-g}
\left\lbrace {1 \over f(\phi)} \left(  \sqrt{1-2f(\phi) X} -1\right) -V(\phi)  \right\rbrace
\end{equation}
where $g$ is determinant of background metric, assumed to be spatially flat FLRW metric, $f(\phi)$ is the brane tension, $V(\phi)$ is an arbitrary potential, and the kinetic term of the scalar field is indicated by $X=-(1/2)g^{\mu\nu}\nabla_\mu\phi \nabla_\nu\phi$. The Firedmann equations of the model are read as
\begin{equation}\label{Friedmann}
H^{2}={1 \over 3M_{p}^{2}}\; \rho_\phi, \qquad \dot{H}={-1 \over 2M_{P}^{2}}(\rho_{\phi} + p_{\phi})
\end{equation}
in which $\rho_\phi$ and $p_\phi$ are respectively the energy density and pressure of DBI scalar field, as
\begin{equation}\label{rho-p}
\rho={(\gamma-1) \over f(\phi)} + V(\phi), \qquad p={\gamma-1 \over \gamma f(\phi)} - V(\phi);
\end{equation}
where
\begin{equation*}
\gamma \equiv {1 \over \sqrt{1-2f(\phi)X} }.
\end{equation*}
The scalar field equation of motion is given by
\begin{equation}
\ddot\phi + {3H \over \gamma^2} \; \dot\phi + {V' \over \gamma^3} + {f'(\gamma+2)(\gamma-1) \over 2f\gamma(\gamma+1)}\; \dot\phi^2 = 0.
\end{equation}
where dot denotes derivative with respect the cosmic time, $t$, and prime indicates derivative with respect the scalar field ,$\phi$. Above equations are the main model dynamical equations that we need to proceed our work in the rest of the paper.\\

%=============================================================%
%=============================================================%
%============== Section 2 =======================================%
%=============================================================%
%=============================================================%
\section{Intermediate Inflation and Perturbation}
\label{Sec3}
As it was mentioned, one way to study inflation is introducing a scale factor. Intermediate inflation is the most interesting one that belongs to this class. In this case, the scale factor is given by $a(t)=a_0 \exp(A t^{\alpha})$, where $0<\alpha<1$, describing an accelerated expansion between power-law and de-Sitter expansion. Consequently, the Hubble parameter and its time derivative is obtained
\begin{equation}\label{Friedmann01}
H = A \alpha t^{\alpha-1}, \qquad  \dot{H} = -A\alpha(1-\alpha)t^{\alpha-2}.
\end{equation}
Substituting the resulted relation for $\dot{H}$ in the Friedmann equation, comes to following expression
\begin{equation}\label{phidot}
\dot\phi^2 = {D^2(t)f(\phi) \over 2} \left[ -1 + \sqrt{1 + {4 \over D^2(t)f^2(\phi)}}\; \right],
\end{equation}
where the parameter $D(t)$ is defined as
\begin{equation}
D(t) \equiv 2 M_p^2 A \alpha (1-\alpha)\; t^{\alpha-2}.
\end{equation}
Then, the scalar field potential could be extracted from the Friedmann equation (\ref{Friedmann}) and Eq.(\ref{rho-p}) as
\begin{equation}\label{potential}
V(\phi) = 3M_p^2 H^2 - {(\gamma - 1) \over f(\phi)}.
\end{equation}
Note that, from Eq.(\ref{phidot}), one could derive the scalar field in term of time. Therefore, the potential could be expressed in term of scalar field which is more desirable since we are more interested to consider the potential behavior versus scalar field. \\
The slow-rolling parameters for the model are introduced by \cite{Bessada,Spalinski}
\begin{eqnarray}\label{srp}
\varepsilon_{H} & = & -{1\over H}\;{d\ln(H) \over dt}, \nonumber \\
\eta_{H} & = & -{1\over H}\; {d\ln(\varepsilon) \over dt}, \nonumber \\
\sigma_{H} & = & -{1\over H}\; {d\ln(c_s) \over dt}.
\end{eqnarray}
where $c_s=\gamma^{-1}$ is the sound speed. The slow-rolling parameters are sometimes stated in terms of number of e-folds $N$ \cite{Bessada}, indicating the amount of inflation, given by
\begin{equation}\label{efold}
N=\int H dt = A(t_e^\alpha - t_i^\alpha).
\end{equation}

The biggest achievement of slow-rolling inflation is its consistency with observational data. Two important predictions of inflation is scalar and tensor perturbation. WMAP and Planck are our latest observational data which brings almost a good insight about parameters associated with these kinds of perturbations. \\
The perturbation topic will be briefly reviewed and the readers could refer to \cite{Garriga} where the perturbation in k-inflation has been studied for the first time. The evolution equation of quantum perturbation parameters $v_k$ is derived as \cite{Tzirakis}
\begin{equation}\label{vk}
v''_k + \left( c_s^2 k^2 - {z'' \over z} \right) v_k=0
\end{equation}
where prime denotes derivative with respect to the conformal time $\tau = \int dt/a$, and $k$ stands for the wave number. The equation is the same equation that is derived for the classical model of inflation with the exception that the parameter $z$ has a different definition as
\begin{equation*}
z = {a\sqrt{\rho+p} \over c_s H} = {a\gamma^{3/2}\dot\phi \over H}
\end{equation*}
Changing the conformal time variable as $x=k/\gamma a H$, and rewriting the equation up to first order of the slow-rolling parameters result in
\begin{equation}\label{vkdiff}
 (1-2\epsilon-2\sigma) {d^2 v_k \over dx^2} - {\sigma \over x } {d v_k \over dx} + \Big[ 1 - {2(1+\epsilon-{3 \over 2}\eta - {3 \over 4}\sigma) \over x^2} \Big]v_k=0
\end{equation}
The solution could be expressed in term of Hackle function as \cite{Tzirakis}
\begin{equation}\label{vksol}
v_k(x) = {1 \over 2} \sqrt{{\pi \over c_s k} \; {x \over 1- \epsilon - \sigma}} \; H_\nu\Big( {x \over 1- \epsilon - \sigma} \Big)
\end{equation}
where $\nu={3\over 2}+2\epsilon-\eta+\sigma$. Scalar and tensor amplitude of perturbation are given respectively by \cite{Bessada},

\begin{equation}\label{scalaramplitude}
\mathcal{P}_s = {H^2 \over 8\pi^2 M_p^2 c_s \varepsilon}, \qquad \mathcal{P}_t = {2H^2 \over \pi^2 M_p^2}.
\end{equation}
Consequently, the scalar and tensor spectra indices is derived in terms of slow-roll parameters as
\begin{eqnarray}
n_s-1  & = &  {d\ln(\mathcal{P}_s) \over d\ln(k)}=-2\varepsilon_H + \eta_H + \sigma_H, \nonumber \\
n_t  & = &  {d\ln(\mathcal{P}_t) \over d\ln(k)}=-2\varepsilon_H .
\end{eqnarray}
The other important perturbation parameter is tensor-to-scalar ratio, which is applied to indirectly measure the tensor perturbation, shown as
\begin{equation}
r = {\mathcal{P}_t \over \mathcal{P}_s} = 16c_s\varepsilon
\end{equation}

%=============================================================%
%=============================================================%
%============== Section 2 =======================================%
%=============================================================%
%=============================================================%
\section{Inflationary era and Observational data}
\label{Sec4}
During this section, we are going to investigate the resulted equations in more detail, obtain the perturbation parameters of the model and compare them with observational data. To go further, the function $f(\phi)$ is required to be specified as a function of scalar field in order to gain the scalar field in terms of time. Since finding the exact solution needs to bring out a specific form of the function $f(\phi)$, we postpone it to the subsequent subsections, where two typical example for the function will be examined.\\
The inflationary era is a period of the universe evolution that it stands is a positive accelerated expansion, that is, the slow-roll parameter $\varepsilon$, expressed by $\varepsilon_H = (1-\alpha) / \alpha\tilde{A}\; T^\alpha$, (where $T=M_pt$ is a dimensionless parameter of time, and $\tilde{A}$ is redefined dimensionless constant, $\tilde{A}=M^{-\alpha}A$) should be smaller than unity. To have an accelerated expansion phase the condition $T^\alpha > (1-\alpha)/\alpha\tilde{A}$ should be satisfied, namely $\varepsilon_H<1$. It is supposed that inflation occurs at earliest possible time, nearly at $T_i^\alpha=(1-\alpha)/\alpha\tilde{A}$. Then, utilizing Eq.(\ref{efold}), the whole important parameters $n_s$, $r$, and amplitude of perturbations could be rewritten in term of number of e-fold. \\
From Planck data, the amplitude of scalar perturbation is about $\ln\Big(10^{10}\mathcal{P}^2_s \Big) = 3.094 \pm 0.034$ (Planck TT,TE,EE+lowP), and the scalar spectra index, which is equal to one for a scale invariant spectrum, is measured about $n_s = 0.9645 \pm 0.0049$ ($68\%$ CL, Planck TT,TE,EE+lowP) \cite{planck2015}. In contrast with scalar perturbation, Planck does not give an exact value for tensor-to-scalar ratio $r$; it just specifies an upper bound for this parameter as $r < 0.10$ ($95\%$ CL, Planck TT,TE,EE+lowP) \cite{planck2015}. \\
To check whether the model could be account as an acceptable model explaining the mechanism of inflation, the perturbation parameters of the model should be compared with the data. However, as we check the parameters it could be understood that we need to know the scalar field as a function of time. The equation for a general inverse power-law function of $f(\phi)$ (namely $f(\phi)=\lambda \phi^{-n}$ where the constant $\lambda$ has dimension $M^{4-n}$) is rewritten as
\begin{equation}
\dot\Phi^2_{,T}(T) = {\tilde{D}^2(T)\tilde{\lambda} \over 2 \Phi^n(T)} \Bigg[ -1 + \sqrt{ 1 + {\Phi^{2n}(T) \over \tilde{D}^2(T)\tilde{\lambda}^2} } \; \Bigg]
\end{equation}
where $\Phi$ is the dimensionless scalar field, $\Phi \equiv \phi / M_p$, $\tilde{\lambda}$ is the dimensionless definition of $\lambda$, $\tilde{D}(T)$ is defined by $\tilde{D}(T)=2\tilde{A}\alpha(1-\alpha)T^{\alpha-2}$, and "$_{,T}$" is $d/dT$. It is necessary to solve the above differential equation for $\phi(t)$ in order to compute the parameter as well as considering the potential behavior during the inflationary era. Unfortunately, getting analytical solution is beset by some difficulties and we are left with the numerical approach. In two following subsections, the model is considered for two cases, and the final result will be compared with latest observational data. \\

%============================================================%
%=============================================================%
%=============================================================%
\subsection{First case: $n=4$}
\begin{figure}
\centering
\subfigure[scalar field versus time]{\includegraphics[width=7cm]{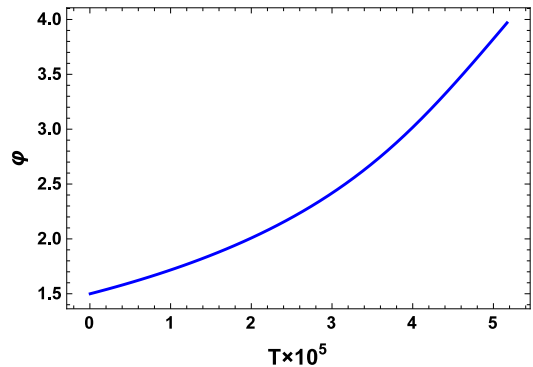}}
\subfigure[potential versus scalar field]{\includegraphics[width=7cm]{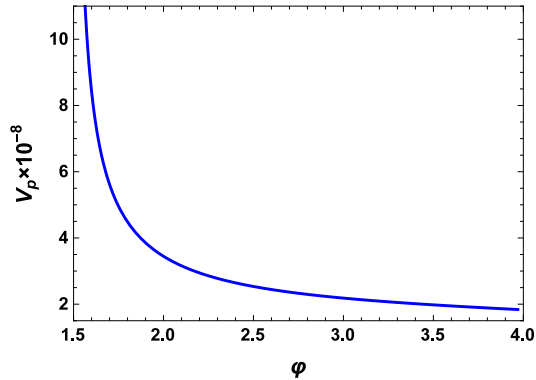}}
\caption{\footnotesize \textbf{a)} the scalar field versus time. \textbf{b)} the potential versus scalar field. The constant parameters have been picked out as: $\alpha=0.67$, $\tilde{A}=8.99 \times 10^{-3}$, $\tilde{\lambda}=1.4\times 10^{12}$ and $N=60$. The parameters $\varphi$ and $V_p$ are respectively defined by the dimensionless scalar field given by $\varphi \equiv \phi/M_p$ and the dimensionless  potential defined by  $V_p \equiv V/M_p^4$.}
\label{phipotI}
\end{figure}

\begin{figure}
\centering
\includegraphics[width=7cm]{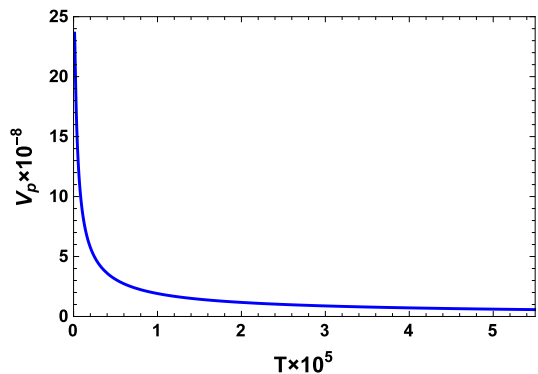}
\caption{\footnotesize The potential versus time for the first case. The constant parameters have been picked out as: $\alpha=0.67$, $\tilde{A}=8.99 \times 10^{-3}$, $\tilde{\lambda}=1.4\times 10^{12}$ and $N=60$. }
\label{pott01}
\end{figure}

The first thing we should deal with is the behavior of the scalar field in terms of time, namely solving the differential Eq.(\ref{phidot}). The differential equation is solved numerically and the result displayed in Fig.\ref{phipotI}. As it could be realized, the scalar field is larger than Planck mass, and grows by increasing time. The potential of the case could be studied by employing the numerical result about the scalar field in the general equation of the potential (\ref{potential}). In Fig.\ref{phipotI}, the potential during the inflationary era is drown. It is clearly seen that the potential at the beginning of inflation is much smaller than Planck mass and it decreases by passing time. The situation is same as Tachyon scalar field potential where it is used to describe inflation in \cite{aghamohammadi}. Obtaining a right potential function needs numerical fit. However, approximately, by considering Eq.(\ref{potential}), it could be stated that the potential is a polynomial function. Exploring the inflationary potential for different types of models, it could be deduced that our potential generally is close to an inverse power-law potential, or an exponential potential with negative argument. \\
%%%%%%%%%%%%%%%%%%%%%%%%%%%%%%%%%%%%%%%%%%%%%%%%%%%%%%%%%%%
%\begin{widetext}
%\begin{center}
%\end{center}
%\end{widetext}
%%%%%%%%%%%%%%%%%%%%%%%%%%%%%%%%%%%%%%%%%%%%%%%%%%%%%%%%%%%
On the other hand, one of the unsolved problem of the inflationary cosmology is the energy scale of the phenomenon. Utilizing the exact derived equations (\ref{potential}) and (\ref{srp}), we are going to take a brief look at the subject. From Eq.(\ref{srp}), one realizes that the slow-roll parameter $\epsilon$ is decreasing with time, then the case $\epsilon = 1$ is associated with the initial time of inflation, so that there is
\begin{equation}\label{initial-time}
\epsilon = 1 = {1-\alpha \over \tilde{A} \alpha T^\alpha} , \qquad T_i^\alpha = {1-\alpha \over \tilde{A} \alpha}
\end{equation}
Taking $\alpha=0.64$, $\tilde{A}=7.99 \times 10^{-3}$, $\tilde{\lambda}=6.68 \times 10^{12}$ and $N=65$, the beginning time of inflation is determined about $T_i=7.7\times 10^2$, in another word $t_i=2.08\times 10^{-40} \rm{s}$. Substituting the time in Eq.(\ref{potential}), the potential at the start point of inflation is obtained about $V_i=2.16 \times 10^{-2}M_p \simeq 5.26 \times 10^{16} \rm{GeV}$, that is smaller than Planck energy scale which in turn shows that we stand in the classical regime and the quantum effect could be ignored. However, the most important energy scale we want to know is the energy scale at the time of perturbation horizon exit, where the whole observational data we detect correspond to that point. The energy scale of the universe at the horizon exit is less that the beginning one, that is approximated by $7.45\times 10^{-3}M_p \simeq 1.81 \times 10^{16} \rm{GeV}$. Briefly, in the presented model, inflation begins at an energy scale smaller than the Planck energy scale and then it reduces so that the energy scale of the universe arrives at $\simeq 10^{16} \rm{GeV}$ where the slow-rolling approximation is perfectly satisfied and the perturbations cross the horizon. Figures.\ref{pott01} illustrates the fact clearly. \\
%%%%%%%%%%%%%%%%%%%%%%%%%%%%%%%%%%%%%%%%%%%%%%%%%%%%%%%%%%%
%%%%%%%%%%%%%%%%%%%%%%%%%%%%%%%%%%%%%%%%%%%%%%%%%%%%%%%%%%%
%%%%%%%%%%%%%%%%%%%%%%%%%%%%%%%%%%%%%%%%%%%%%%%%%%%%%%%%%%%

\begin{figure}
\centering
\subfigure[$r-n_s$]{\includegraphics[width=7cm]{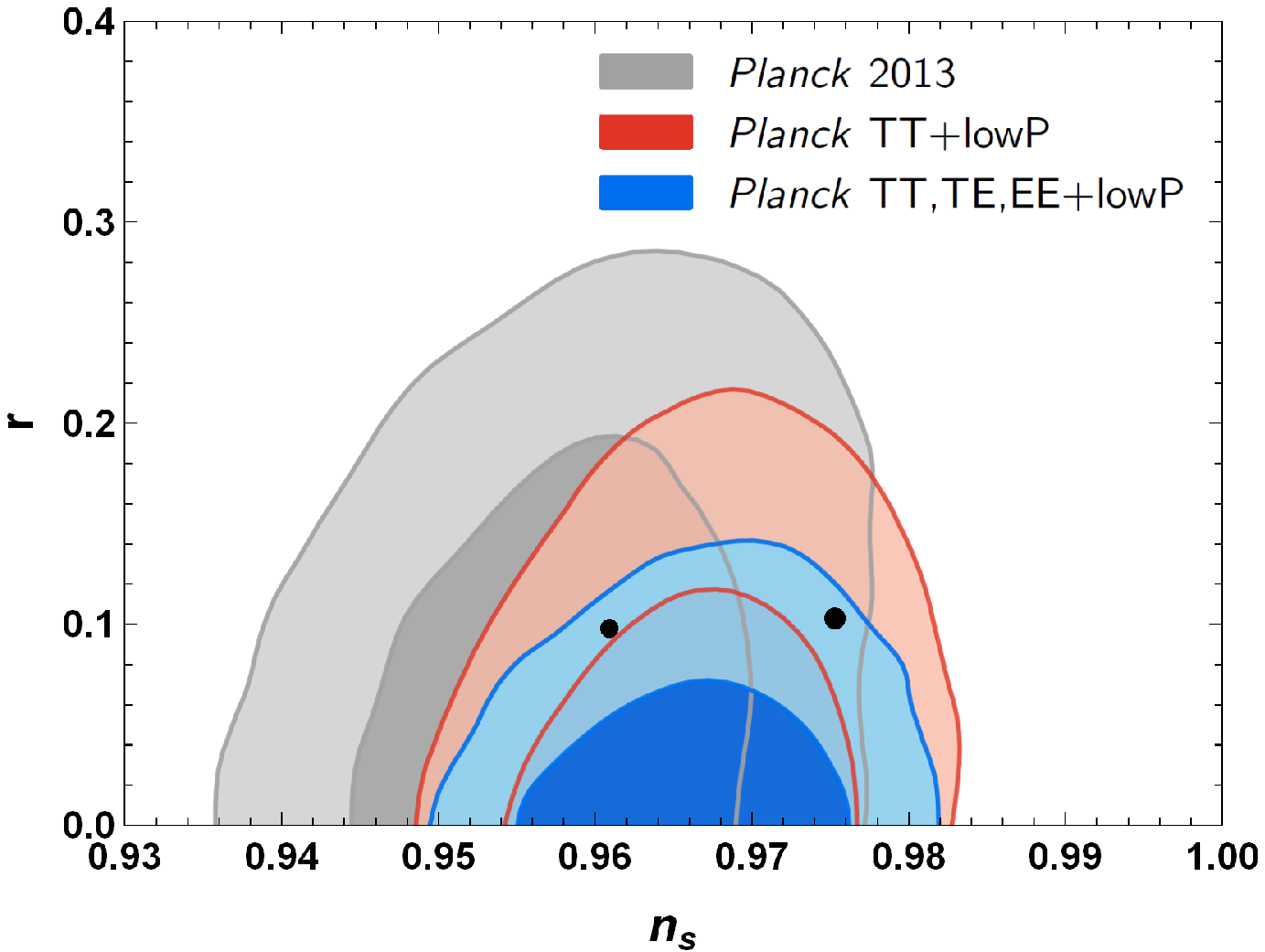}}
\subfigure[${dn_s \over dlnk}_s$]{\includegraphics[width=7cm]{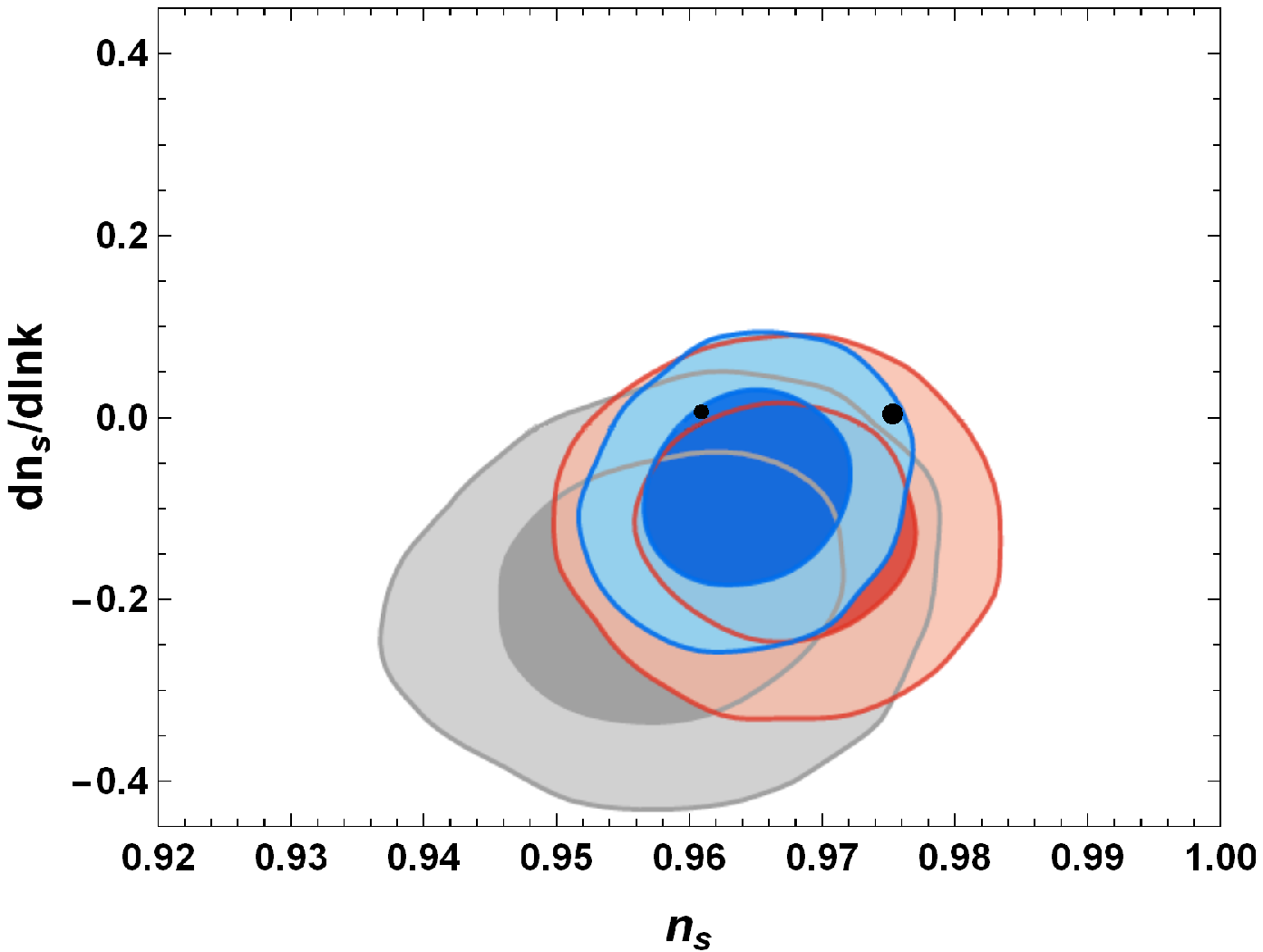}}
  \caption{\footnotesize \textbf{a)} The $r-n_s$ diagram; \textbf{b)} ${dn_s \over dlnk}-n_s$ : for free parameters: $\alpha=0.67$, $\tilde{A}=8.99 \times 10^{-3}$, $\tilde{\lambda}=1.4\times 10^{12}$. The middle and large points are respectively related to $N=60$ and $N=63$. }\label{Fnsr}
\end{figure}

\noindent It might be right to say that the most important result of Planck data is $r-n_s$ diagram. It is an appropriate way to categorize the inflationary models based on their prediction about $r$ and $n_s$, in which the more reliable models are those that have the best prediction for the diagram. Thus, we are going to concentrate on this diagram for the model. The final result is illustrated in Fig.\ref{Fnsr}, which stands in $95\%$ CL area of Planck data. Then, it could be concluded that the case could still be considered as a valid candidate for explaining the inflationary scenario. The running scalar spectra index is the next step to test the model. The model prediction about the parameter is presented in Fig.\ref{Fnsr}, which displays that the result stay in $95\%$ CL, however the point related to $N=60$ stays in the boundary era between $68\%$ CL and $95\%$ CL regions.   \\

%%%%%%%%%%%%%%%%%%%%%%%%%%%%%%%%%%%%%%%%%%%%%%%%%%%%%%%%%%%
%\begin{widetext}
%\begin{center}
%\end{center}
%\end{widetext}
%%%%%%%%%%%%%%%%%%%%%%%%%%%%%%%%%%%%%%%%%%%%%%%%%%%%%%%%%%%

\noindent In Table.\ref{FT}, one can see the model prediction for perturbation parameters by other choice of free parameters. It is realized that the parameters stand in almost suitable range and it could be concluded that the model has the ability to stand as a suitable case for portraying inflation. Note that, according to Eq.(\ref{scalaramplitude}) the amplitude of scalar perturbation $\mathcal{P}_s$ is generally depends on $a(t)$. However, the presented $\mathcal{P}_s$ in the above table, is the amplitude of scalar perturbation at the time of horizon exit. This parameter is same as $A_s$ in Planck papers \cite{planck2015}, which expresses $\mathcal{P}_s$ at $k=k_\ast$.  \\
%%%%%%%%%%%%%%%%%%%%%%%%%%%%%%%%%%%%%%%%%%%%%%%%%%%%%%%%%%%%%%%%%%%%%%%%%%
%%%%%%%%%%%%%%%%%%%%%%%%%%%%%%%%%%%%%%%%%%%%%%%%%%%%%%%%%%%%%%%%%%%%%%%%%%
%\begin{widetext}

\begin{table}[h]
  \centering
  {\footnotesize
  \begin{tabular}{p{0.8cm}p{0.8cm}p{0.8cm}p{0.8cm}p{1.2cm}p{1cm}p{1.5cm}}
    \toprule[1.5pt] \\[-2.5mm]
              $\ \ \alpha$   & $\mathcal{A}$ & $\zeta$ & $N$ &  $\ \ n_s$   & $\ \ r$  & $\qquad \mathcal{P}_s$    \\[0.01mm]
      \midrule[1.5pt] \\[-2.5mm]
         $0.64$   & $7.99$ & $6.85$ &  $70$ &  $0.9647$   & $0.094$ & $3.27 \times 10^{-9}$   \\[2mm]
         $0.64$   & $7.99$ & $6.68$ &  $65$ &  $0.9692$   & $0.109$ & $2.30 \times 10^{-9}$   \\[2mm]
         $0.42$  & $9.00$ & $1.30$ &  $60$ &  $0.9686$   & $0.104$ & $1.20 \times 10^{-8}$   \\[0.1mm]
    \bottomrule[1.5pt]
  \end{tabular}
  }
  \caption{\footnotesize The model prediction about the perturbation parameters $n_s$, $r$, $\mathcal{P}_s$ are prepared for different values of the free parameters of the model(the redefined parameters are $\mathcal{A}=\tilde{A} \times 10^3$ and $\zeta=\tilde{\lambda} \times 10^{-12}$).}\label{FT}
\end{table}

%\end{widetext}
%%%%%%%%%%%%%%%%%%%%%%%%%%%%%%%%%%%%%%%%%%%%%%%%%%%%%%%%%%%%%%%%%%%%%%%%%%
%%%%%%%%%%%%%%%%%%%%%%%%%%%%%%%%%%%%%%%%%%%%%%%%%%%%%%%%%%%%%%%%%%%%%%%%%%

%============================================================%
%=============================================================%
%=============================================================%
\subsection{Second case: $n=2$}
\begin{figure}
\centering
\subfigure[scalar field versus time]{\includegraphics[width=7cm]{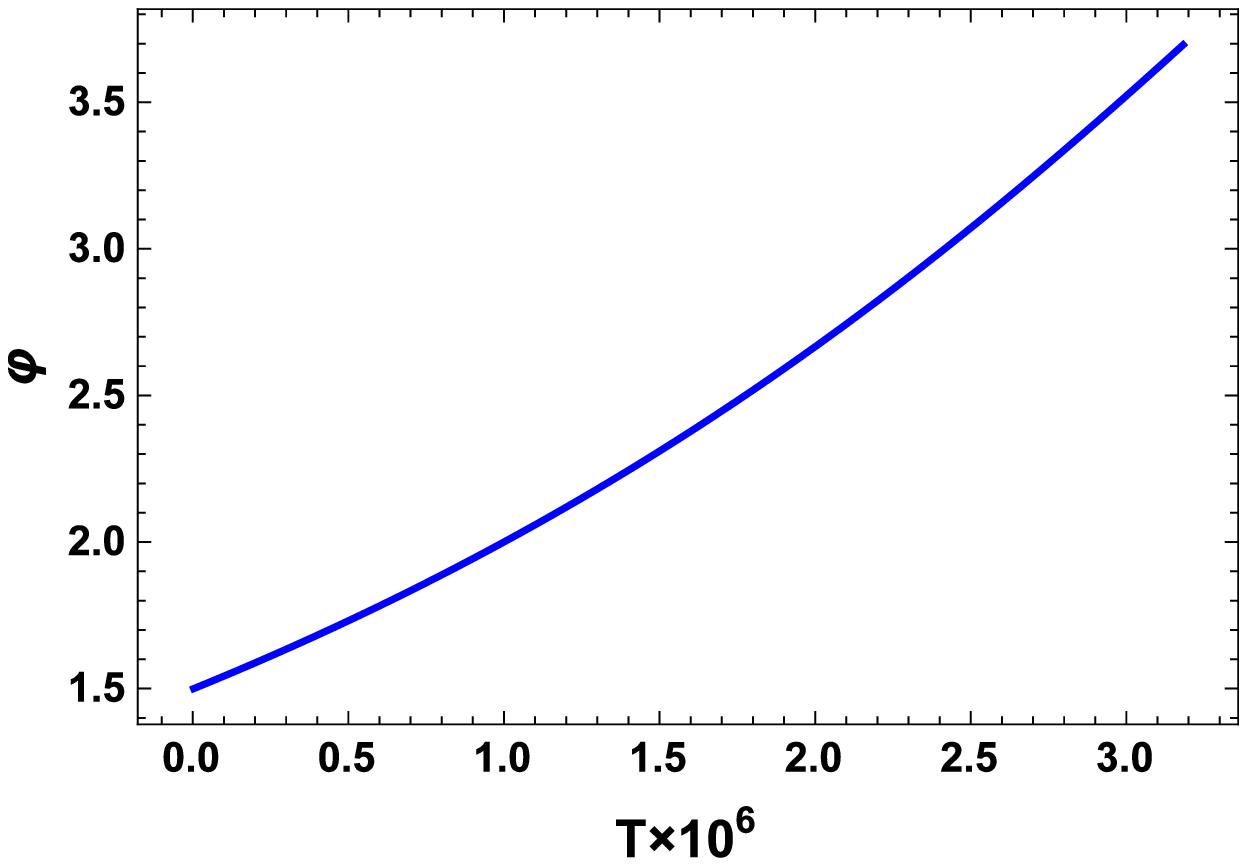}}
\subfigure[potential versus scalar field]{\includegraphics[width=7cm]{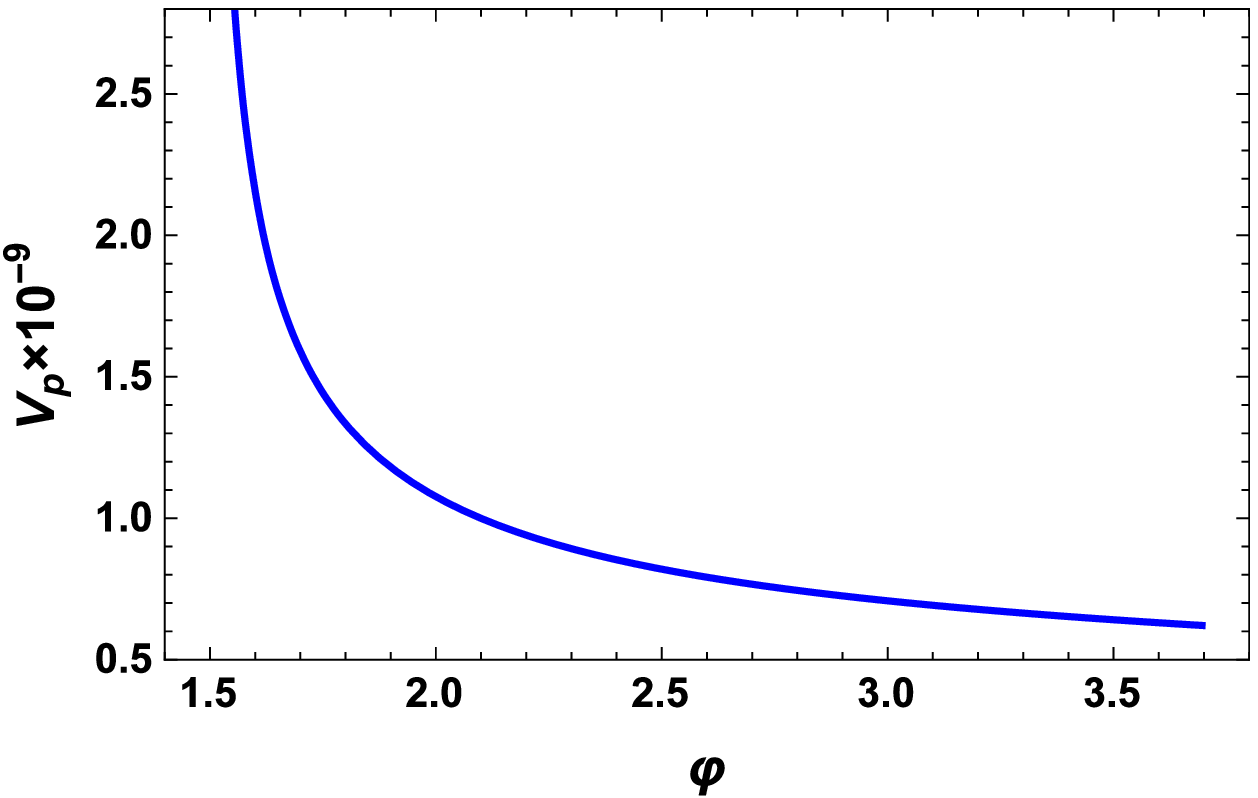}}

\caption{\footnotesize \textbf{a)} the scalar field versus time. \textbf{b)} the potential versus scalar field. The constant parameters have been picked out as: $\alpha=0.76$, $\tilde{A}=6.89 \times 10^{-4}$, $\tilde{\lambda}=1.2\times 10^{13}$ and $N=60$.}
\label{phipotII}
\end{figure}

By solving the differential equation (\ref{phidot}) numerically for case, the scalar field is plotted in Fig.\ref{phipotII}. The potential could be depicted in Fig.\ref{phipotII}, using Eq.(\ref{potential}). It is found out that the potential decreases by passing time and increasing the scalar field. The potential keeps almost the same behavior as previous case; it starts on scale smaller than Planck energy and it reduces by increasing time. The shape of the potential could be approximated as a polynomial function including inverse power-law or exponential terms of 
scalar field. \\
%%%%%%%%%%%%%%%%%%%%%%%%%%%%%%%%%%%%%%%%%%%%%%%%%%%%%%%%%%%
%%%%%%%%%%%%%%%%%%%%%%%%%%%%%%%%%%%%%%%%%%%%%%%%%%%%%%%%%%%
To talk about the energy scale of the inflation in this case, it should be mentioned that almost the same result as the previous case is derived here as well, in which the inflation is starts about $8.18 \times 10^{-38} \rm{s}$ after the big bang that it means for this case inflation begins later that the first case. Besides, the inflation energy scale at horizon exits, is predicted about $10^{16} \rm{GeV}$ that is the same energy scale as the previous case.  \\
%%%%%%%%%%%%%%%%%%%%%%%%%%%%%%%%%%%%%%%%%%%%%%%%%%%%%%%%%%%
\begin{figure}
\centering
\includegraphics[width=7cm]{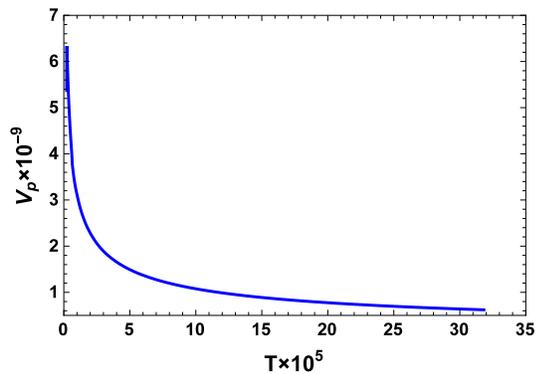}
\caption{\footnotesize The potential versus time for the second case. The constant parameters have been picked out as: $\alpha=0.67$, $\tilde{A}=8.99 \times 10^{-3}$, $\tilde{\lambda}=1.4\times 10^{12}$ and $N=60$. }
\label{pott02}
\end{figure}
%%%%%%%%%%%%%%%%%%%%%%%%%%%%%%%%%%%%%%%%%%%%%%%%%%%%%%%%%%%
%%%%%%%%%%%%%%%%%%%%%%%%%%%%%%%%%%%%%%%%%%%%%%%%%%%%%%%%%%%

\begin{figure}
  \centering
  \subfigure[$r-n_s$]{\includegraphics[width=7cm]{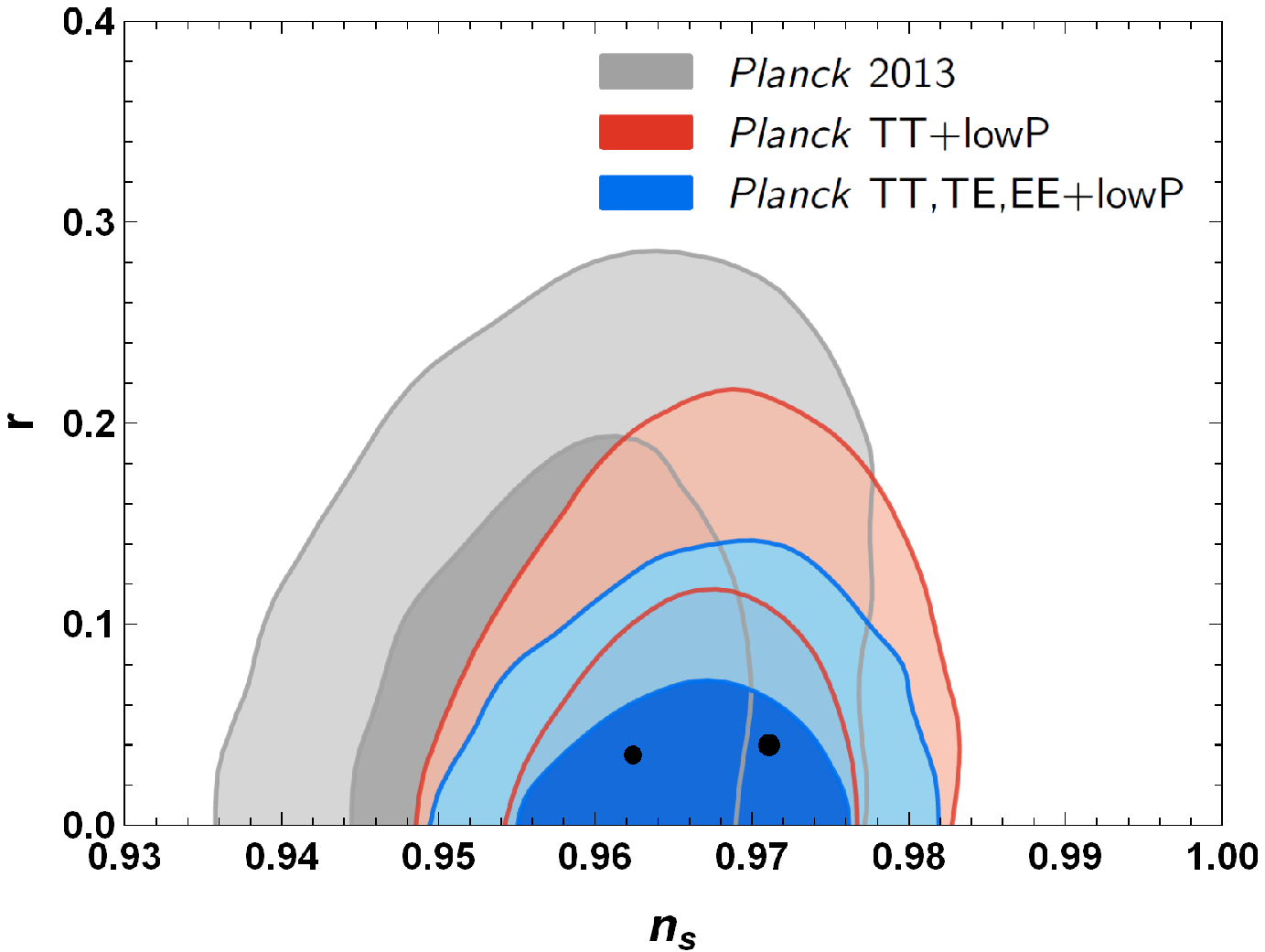}}
\subfigure[${dn_s \over dlnk}_s$]{\includegraphics[width=7cm]{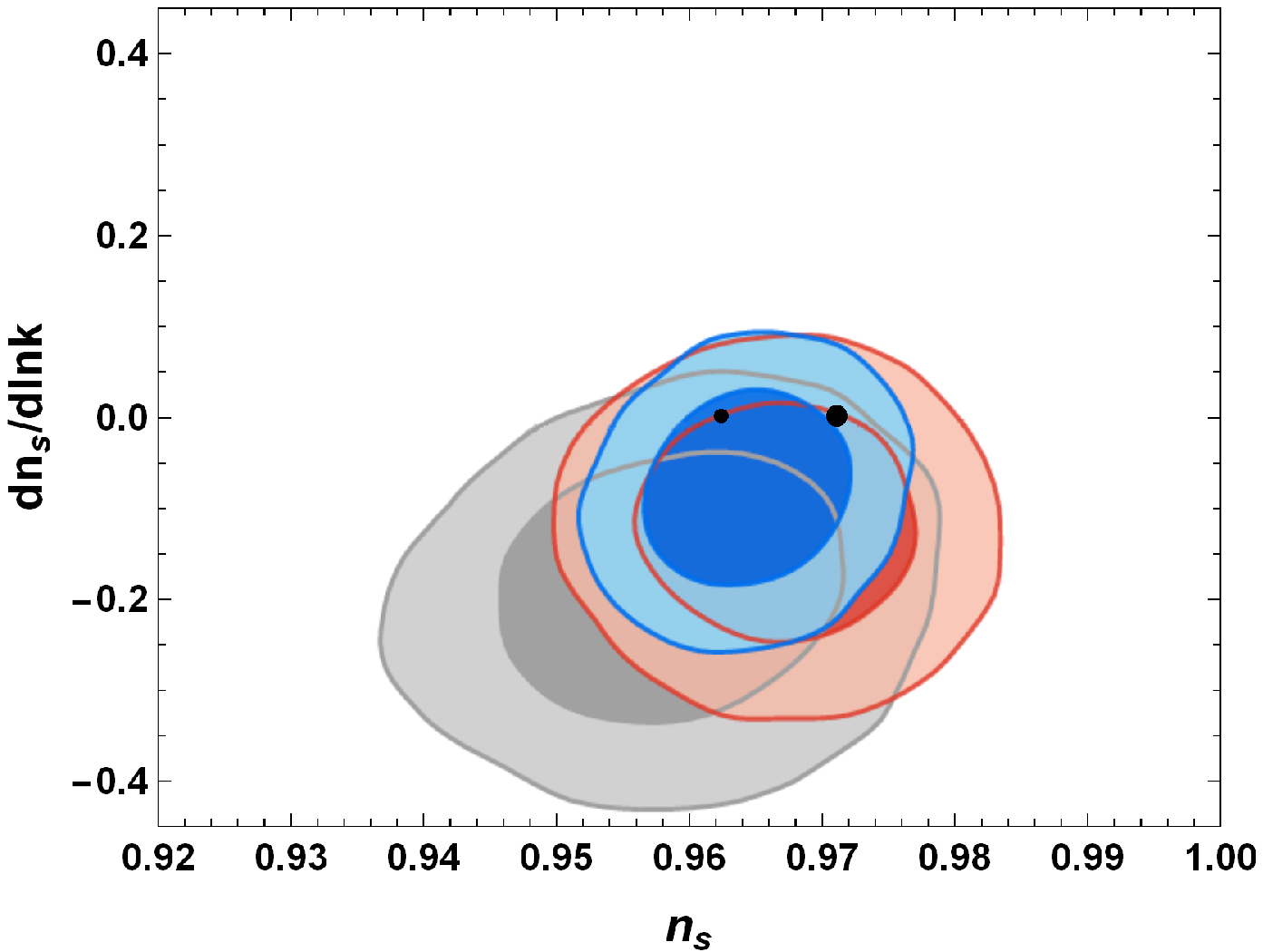}}
  \caption{\footnotesize \textbf{a)} The $r-n_s$ diagram; \textbf{b)} ${dn_s \over dlnk}-n_s$ : for free parameters: $\alpha=0.76$, $\tilde{A}=6.89 \times 10^{-4}$, $\tilde{\lambda}=1.2\times 10^{13}$. The middle and large points are respectively related to $N=60$ and $N=65$.}\label{Snsr}
\end{figure}

\noindent Computing the slow-roll parameters allows one to plot $r-n_s$ diagram, as in Fig.\ref{Snsr}. In a comparison with Planck data, it is realized that the points stay in $68\%$ CL which is a more desirable result than the previous case, and expresses a great consistency with the Planck data. However, the result for running scalar spectra index, as illustrated in Fig.\ref{Snsr}, states that the point related to $N=60$ stands in the $68\%$ CL region, and the other point stays in the boundary between the $68\%$ CL and $95\%$ CL regions.   \\
%%%%%%%%%%%%%%%%%%%%%%%%%%%%%%%%%%%%%%%%%%%%%%%%%%%%%%%%%%%
%%%%%%%%%%%%%%%%%%%%%%%%%%%%%%%%%%%%%%%%%%%%%%%%%%%%%%%%%%%

\noindent The model prediction for perturbation parameters is prepared in Table.\ref{ST}, for different choice of free parameters. It exhibits that the parameters estimated by the model stand on almost suitable range with observational data, so that the parameters $n_s$ and $r$ are in good agreement with Planck data and the scalar perturbation amplitude is smaller by one order of magnitude.   \\

%%%%%%%%%%%%%%%%%%%%%%%%%%%%%%%%%%%%%%%%%%%%%%%%%%%%%%%%%%%%%%%%%%%%%%%%%%
%%%%%%%%%%%%%%%%%%%%%%%%%%%%%%%%%%%%%%%%%%%%%%%%%%%%%%%%%%%%%%%%%%%%%%%%%%
%\begin{widetext}

\begin{table}[h]
  \centering
  {\footnotesize
  \begin{tabular}{p{0.8cm}p{0.8cm}p{0.8cm}p{0.8cm}p{1.2cm}p{1cm}p{1.7cm}}
    \toprule[1.5pt] \\[-2.5mm]
              $\ \ \alpha$   & $\mathcal{A}$ & $\zeta$ & $N$ &  $\ \ n_s$   & $\ \ r$  & $\qquad \mathcal{P}_s$   \\[0.01mm]
      \midrule[1.5pt] \\[-2.5mm]
         $0.72$   & $5.89$ & $1.23$ &  $70$ &  $0.9639$   & $0.040$ & $1.002 \times 10^{-10}$   \\[2mm]
         $0.76$   & $6.89$ & $1.26$ &  $65$ &  $0.9692$   & $0.038$ & $1.047 \times 10^{-9}$   \\[2mm]
         $0.75$   & $4.20$ & $6.00$ &  $60$ &  $0.9647$   & $0.042$ & $1.725 \times 10^{-10}$   \\[0.1mm]
    \bottomrule[1.5pt]
  \end{tabular}
  }
  \caption{\footnotesize The model prediction about the perturbation parameters $n_s$, $r$, $\mathcal{P}_s$ are prepared for different values of the free parameters of the model(the redefined parameters are $\mathcal{A}=\tilde{A} \times 10^4$ and $\zeta=\tilde{\lambda} \times 10^{-13}$.}\label{ST}
\end{table}

%\end{widetext}
%%%%%%%%%%%%%%%%%%%%%%%%%%%%%%%%%%%%%%%%%%%%%%%%%%%%%%%%%%%%%%%%%%%%%%%%%%
%%%%%%%%%%%%%%%%%%%%%%%%%%%%%%%%%%%%%%%%%%%%%%%%%%%%%%%%%%%%%%%%%%%%%%%%%%

%%%%%%%%%%%%%%%%%%%%%%%%%%%%%%%%%%%%%%%%%%%%%%%%%%%%%%%%%%%
%%%%%%%%%%%%%%%%%%%%%%%%%%%%%%%%%%%%%%%%%%%%%%%%%%%%%%%%%%%
%%%%%%%%%%%%%%%%%%%%%%%%%%%%%%%%%%%%%%%%%%%%%%%%%%%%%%%%%%%
%%%%%%%%%%%%%%%%%%%%%%%%%%%%%%%%%%%%%%%%%%%%%%%%%%%%%%%%%%%
%%%%%%%%%%%%%%%%%%%%%%%%%%%%%%%%%%%%%%%%%%%%%%%%%%%%%%%%%%%
%%%%%%%%%%%%%%%%%%%%%%%%%%%%%%%%%%%%%%%%%%%%%%%%%%%%%%%%%%%
%%%%%%%%%%%%%%%%%%%%%%%%%%%%%%%%%%%%%%%%%%%%%%%%%%%%%%%%%%%
%%%%%%%%%%%%%%%%%%%%%%%%%%%%%%%%%%%%%%%%%%%%%%%%%%%%%%%%%%%
\section{Non-Gaussianity}
Further observational constraint could be applied for any inflationary models by studying the non-Gaussianity. The primordial fluctuation could be Gaussian or non-Gaussian. If the primordial fluctuation are Gaussian, then all information are stored in power spectrum, or in another word in two-point correlation function which are derived from action by expanding it to second order of curvature perturbation $\mathcal{R}$. On the other side, the higher order correlation function are required when the fluctuation are non-Gaussian.  There are different shapes for non-Gaussianity that are describe by triangles, so that any inflationary model predicts a maximum non-Gaussianity for a specific shape of non-Gaussianity. Equilateral triangle is a kind of non-Gaussianity that predicts by non-canonical models of inflationary models where the scalar field kinetic term is modified and there is a non-trivial sound speed. Although the parameter has not been exactly determined from Planck data, there is a range for the parameters that should be satisfied for any acceptable model.\\
The magnitude of non-Gaussianity that measured in the observational data is defined as
\begin{equation}
f_{NL} = {5 \over 18} \; {\mathcal{B}_{\mathcal{R}}(k,k,k) \over \mathcal{P}_{\mathcal{R}}^2(k)}
\end{equation}
where $\mathcal{B}$ is three point correlation function, named as bispectrum. The parameter is extracted from the third order of the action. The third order action appears with a slow-rolling factor that reflect the fact that for the slow-rolling inflation the non-Gaussianity is small.\\
The non-Gaussianity for non-canonical models of scalar field peaks at the equilateral shape, given by \cite{Bessada, peiris, Baumann}
\begin{equation}
f_{NL}^{equil} = -{35 \over 108} \; \left( { {1 \over c_s^2} -1} \right) + {5 \over 81} \; \left( { {1 \over c_s^2} - 1 - {2 \over \Sigma}} \right),
\end{equation}
where
\begin{equation*}
\Sigma = { X^2 P_{,XX} + {2 \over 3} X^3 P_{,XXX} \over XP_{X}+2X^2P_{XX} },
\end{equation*}
and $P$ is the DBI scalar field Lagrangian expressed in the action. Therefore, the non-Gaussianity for DBI model could be extracted as \cite{Bessada, peiris, Baumann}
\begin{equation}\label{DBIfnl}
f_{NL}^{DBI} = -{35 \over 108} \; \left( { {1 \over c_s^2} -1} \right).
\end{equation}
It has been computed for both cases, and the final result has been prepared by Table.\ref{ng} for different values of e-folds number. The latest Planck data indicates that the non-Gaussianity for the equilateral triangle is about $f_{NL}^{DBI}=-3.7\pm43$ from T and E ($68\%$ CL) \cite{planck2015}.
%%%%%%%%%%%%%%%%%%%%%%%%%%%%%%%%%%%%%%%%%%%%%%%%%%%%%%%%%%%%%%%%%%%%%%%%%%
\begin{table}[h]
  \centering
  \begin{tabular}{p{2cm}||p{2.5cm}|cc}
     \toprule[1.5pt] \\[-3mm]
                & const. param                    & \quad $N$ \quad & $f_{NL}^{equil}$ \\[0.1mm]
     \midrule[1.5pt] \\[-2.5mm]
                & $\alpha = 0.67$                        & \ $55$ & $-0.680$ \\[2mm]
    first case  & $\tilde{A}=8.99 \times 10^{-3}$        & \ $60$ & $-0.242$ \\[2mm]
                & $\tilde{\lambda}=1.40 \times 10^{12}$  & \ $65$ & $-0.105$ \\[0.5mm]
            \hline \\[-4.5mm]
            \hline \\[-2.5mm]
                & $\alpha = 0.75$                        & \ $55$ & $-0.112$ \\[2mm]
    second case & $\tilde{A}=8.99 \times 10^{-3}$        & \ $60$ & $-0.102$ \\[2mm]
                & $\tilde{\lambda}=1.40 \times 10^{12}$  & \ $65$ & $-0.094$ \\[0.1mm]
     \bottomrule[1.5pt]
  \end{tabular}
  \caption{The predicted non-Gaussianity for both cases.}\label{ng}
\end{table}
%%%%%%%%%%%%%%%%%%%%%%%%%%%%%%%%%%%%%%%%%%%%%%%%%%%%%%%%%%%%%%%%%%%%%%%%%%
Then, One could realize that for the both considered case stand in the $68\% CL$ regime, however they are very small and near zero. It sounds that the model predicts that the primordial perturbation made by intermediate DBI inflation is almost Gaussian.

%%%%%%%%%%%%%%%%%%%%%%%%%%%%%%%%%%%%%%%%%%%%%%%%%%%%%%%%%%%
%%%%%%%%%%%%%%%%%%%%%%%%%%%%%%%%%%%%%%%%%%%%%%%%%%%%%%%%%%%
\section{Discussion and Conclusions}\label{conclusion}
The main case of interest in this work was to consider the intermediate inflation by using DBI scalar field as inflaton. Utilizing the dynamical equations of the presented model, and definition of slow-roll parameters, the DBI slow-rolling inflationary scenario was considered in general. The specific form of arbitrary function $f(\phi)$ was required to be able the final result of the model. Then, the rest of the work was divided into two parts related to $f(\phi)\propto\phi^{-4}$ and $\phi^{-2}$. \\
Solving the differential equation (\ref{phidot}) for the first case expressed that the scalar field is larger than Planck mass and it increases by passing time. Then, the potential was drawn displaying a potential smaller than Planck energy density that decreases during the inflationary era. Examining the perturbation parameters was desirable for the case as well. It was found out that the model is able to produce a perfect value for the scalar spectra index which stands in an interval presented by observational data. In addition, the parameter $r$ provided by the model is consistent with the planck data. These result was gathered and illustrated in Fig.\ref{Fnsr} presenting the points in the light blue color area ($95\%$ CL). Thus, it could be figured out that the result given by the model are acceptable in comparison with the Planck data. On the other hand, the running scalar spectra index seems to be at the $95\%$ CL region, however the point related to $N=60$ still could stays in the $68\%$ CL area. It came more interesting when the model estimated the amplitude of scalar perturbation in the same order of magnitude as Planck data; given by Table.\ref{FT}. Considering the starting point of the energy scale of the model is derived smaller than the Planck energy scale, and ensures that the quantum effects on the dynamical equation could be ignored. Besides, the energy scale of the universe when the perturbation exit the horizon was about $10^{16} \rm{GeV}$ for both cases that is in agreement with our believe that inflation occurs at enormous high energy scale.   \\
The same process was repeated in the second case. Solving the differential equation (\ref{phidot}) revealed the scalar field behavior during the inflationary time period, indicating an scalar field larger than Planck mass. Plugging the result in the potential equation (\ref{potential}), make it possible to easily plot it, expressing a potential smaller than Planck energy density which decreases by going forward in time. However, the result of $r-n_s$ diagram for this case was more desirable than the previous one, so that the acquired point stands in the dark blue area ($68\%$ CL), Fig.\ref{Snsr}, pointing out a great consistency with the Planck data. As a further discussion, the result of the model about the running scalar spectra index was depicted as well, describing a better situation than the previous case, in which they stand in $68\%$ CL except for $N=65$ that is very close the mentioned area. At last, the amplitude of the scalar perturbation for the case was estimated and provided in Table.\ref{ST}, that are almost the same magnitude as the Planck data. In addition, for this case, the inflation is predicted to begin about $10^{-38} \rm{s}$ after the big bang, and the energy scale is estimated as same as the previous case, namely about $10^{16} \rm{GeV}$, at horizon exit.  \\
To appreciate the importance of the presented model, let compare the result of this model with \cite{BarowLiddle02}, where intermediate inflation is studied using a canonical scalar field. It should first mentioned that, the initial time of inflation for all intermediate inflation model has the same relation because of the proposed scale factor form. However, the initial scalar field and also the scalar field at horizon exit are different from model to another since the dynamical equation for different models of scalar field are different. Consequently the energy scale of the inflation could be different.  It is clearly seen that for canonical scalar field, the perturbation parameters such as $n_s$ and $r$ are not in acceptable range predicted by Planck. for a value of $\alpha$, if one wants to get a proper value of $r$, then the scalar spectra index becomes larger than unity, and if one wants to get a suitable value of $n_s$, the tensor-to-scalar ratio becomes larger than $0.10$. The problem also could not be fixed by changing the $\alpha$ parameter. Therefore, it seems that the model should be abandoned. However, using a non-canonical scalar field could be a way to solve the problem properly. DBI scalar field, as a model of non-canonical scalar field, was introduced in the work. The obtained result show that, one could get an acceptable value for the important perturbation parameters $n_s$ and $r$, in which they stand in the Planck data range.  \\
As a further argument, the non-Gaussianity prediction of the model was considered in the last section. Since there is a non-canonical model of scalar field assumed for describing inflation with a modified kinetic term, amongst different shapes of non-Gaussianity it has a peak for the equilateral triangle shape. The obtained result states that the non-Gaussianity for two cases stand in $68\%$ CL region, but they are small so that the model predict almost Gaussian perturbation.

%=============================================================%
%=============================================================%
%=======================  References =========================%
%=============================================================%
%=============================================================%



%% References


%\bibliography{sample.bib}



%%%%%%%%%%%%%%%%%%%%%%%%%%%%%%%%%%%%%%%%%%%%%%%%%%%%%%%%%%%%%%%%%%%%%
%%%%%%%%%%%%%%%%%%%%%%%%%%%%%%%%%%%%%%%%%%%%%%%%%%%%%%%%%%%%%%%%%%%%

\end{document}